\documentclass[a4paper,11pt]{article}
\usepackage{pos}

\title{J2102+6015: an Intriguing Radio-loud Active Galactic Nucleus in the Early Universe}
\ShortTitle{J2102+6015: an Intriguing Radio-loud AGN in the Early Universe}

\author*[a,b]{S\'andor~Frey}
\author[c]{Tao~An}
\author[d,e,a]{Krisztina~Gab\'anyi}
\author[f,g]{Leonid~Gurvits}
\author[d,a]{M\'at\'e~Krezinger}
\author[h]{Alexey~Melnikov}
\author[c]{Prashanth~Mohan}
\author[f]{Zsolt~Paragi}
\author[a]{Krisztina~Perger}
\author[c]{Fengchun~Shu}
\author[i]{Oleg~Titov}
\author[j]{Pablo~de~Vicente}
\author[c]{Yingkang~Zhang}

\affiliation[a]{Konkoly Observatory, ELKH Research Centre for Astronomy and Earth Sciences (MTA Centre of Excellence), Konkoly Thege Mikl\'os \'ut 15-17, H-1121, Budapest, Hungary}

\affiliation[b]{Institute of Physics, ELTE E\"otv\"os Lor\'and University, P\'azm\'any P\'eter s\'et\'any 1/A, H-1117, Budapest, Hungary}

\affiliation[c]{Shanghai Astronomical Observatory, Chinese Academy of Sciences, 80 Nandan Road, Shanghai 200030, China}

\affiliation[d]{Department of Astronomy, Institute of Geography and Earth Sciences, ELTE E\"otv\"os Lor\'and University,  P\'azm\'any P\'eter s\'et\'any 1/A, H-1117, Budapest, Hungary}

\affiliation[e]{ELKH-ELTE Extragalactic Astrophysics Research Group, E\"otv\"os Lor\'and University,  P\'azm\'any P\'eter s\'et\'any 1/A, H-1117, Budapest, Hungary}

\affiliation[f]{Joint Institute for VLBI ERIC, Oude Hoogeveensedijk 4, 7991 PD Dwingeloo, The~Netherlands}

\affiliation[g]{Faculty of Aerospace Engineering, Delft University of Technology, Kluyverweg 1, 2629 HS Delft, The Netherlands}

\affiliation[h]{Institute of Applied Astronomy, Russian Academy of Sciences, Kutuzova Embankment 10, St. Petersburg 191187, Russia}

\affiliation[i]{Geoscience Australia, PO Box 378, Canberra 2601, Australia}

\affiliation[j]{Observatorio de Yebes (IGN), Apartado 148, 19180 Yebes, Spain}

\emailAdd{frey.sandor@csfk.org}
\emailAdd{antao@shao.ac.cn}
\emailAdd{k.gabanyi@astro.elte.hu}
\emailAdd{lgurvits@jive.eu}
\emailAdd{krezinger.mate@csfk.org}
\emailAdd{aem@iaaras.ru}
\emailAdd{pmohan@shao.ac.cn}
\emailAdd{zparagi@jive.eu}
\emailAdd{perger.krisztina@csfk.org}
\emailAdd{sfc@shao.ac.cn}
\emailAdd{Oleg.Titov@ga.gov.au}
\emailAdd{pablo.devicente@oan.es}
\emailAdd{ykzhang@shao.ac.cn}

\abstract{The powerful high-redshift quasar J2102+6015 (at $z=4.575$) may provide useful information for studying supermassive black hole growth, galaxy evolution and feedback in the early Universe. The source has so far been imaged with very long baseline interferometry (VLBI) at 2/8~GHz ($S/X$) bands only, showing complex compact structure. Its total radio spectrum peaks at $\sim 6$~GHz in the rest frame. There is no sign of Doppler-boosted jet emission, and the separation of the two major features in its east--west oriented structure spanning $\sim 10$ milliarcsec does not change significantly on a timescale longer than a decade. However, VLBI astrometric monitoring observations suggest quasi-periodic ($\sim 3$~yr) variation in its absolute position. J2102+6015 is presumably a young radio source with jets misaligned with respect to the line of sight. Here we briefly report on our new high-resolution imaging observations made with the European VLBI Network (EVN) at 5 and 22~GHz frequencies in 2021 June, and give an overview of what is currently known about this peculiar distant jetted active galactic nucleus.}

\FullConference{%
  15th European VLBI Network Mini-Symposium and Users' Meeting (EVN2022)\\
  11-15 July 2022\\
  University College Cork, Ireland
}

\begin{document}
\maketitle

\section{Introduction}

Luminous radio-loud quasars in the early Universe (at redshifts $z>4.5$) are extremely rare: only a handful of them are known with monochromatic radio powers reaching the order of $P \sim 10^{28}$~W\,Hz$^{-1}$ at GHz frequencies \cite{2016MNRAS.463.3260C}. They may offer clues for understanding the whole population of high-redshift radio-loud active galactic nuclei (AGN), which can in turn provide constraints on black hole growth, galaxy evolution and feedback in the early Universe, and ways for testing cosmological models. For example, a recent multi-epoch study of J0906+6930 ($z=5.47$) using very long baseline interferometry (VLBI) revealed an evolving pc-scale radio structure and a jet bending likely due to interaction with the dense medium in the central region of the host galaxy \cite{2020NatCo..11..143A}.

The subject of this paper, the radio source J2102+6015 (also known as 2101+600) is another powerful ($P \approx 10^{28}$~W\,Hz$^{-1}$ \cite{2016MNRAS.463.3260C,2018A&A...618A..68F}) radio quasar which has so far been imaged only at $S$ and $X$ bands (around 2 and 8~GHz frequencies) with VLBI \cite{2002ApJS..141...13B,2008AJ....136..580P,2018A&A...618A..68F,2021MNRAS.507.3736Z,2022ApJ...937...19Z}. Its spectroscopic redshift, $z=4.5749$ was measured by \cite{2004ApJ...609..564S}. However, because of the low signal-to-noise ratio in the optical spectrum, the authors labeled this value as marginal. Indeed, the object is rather faint in the optical ($g = 24.9$ and $r = 22.6$ \cite{2016arXiv161205560C}) but the $g-r$ colour difference is not in contradiction with the high redshift \cite{Titov-AJ}. In the absence of  any other independent spectroscopic redshift determination known to us, we shall assume that the value published by \cite{2004ApJ...609..564S} is correct. The quasar is also detected in X-rays ($0.5-7.0$~keV energy band) with the \textit{Chandra} space telescope \cite{2020ApJ...899..127S}, along with 14 other $z>4.5$ radio quasars. 

Prior to 2017, there were only two archival VLBI imaging observations of J2102+6015 at $S$ and $X$ bands, one in 1994 \cite{2002ApJS..141...13B} and another in 2006 \cite{2008AJ....136..580P}. However, there have been a series of recent VLBI observations of this quasar 
(see \cite{2021MNRAS.507.3736Z} for a census of $S/X$-band experiments). For example, sensitive 2.3- and 8.6-GHz images were made in 2017 by combining data taken at 5 different epochs in an astrometric observing programme, with a globally distributed 5-element radio telescope network \cite{2018A&A...618A..68F}. At 2.3~GHz, there was a single slightly resolved component elongated in the east--west direction. At 8.6~GHz, this structure within the central $\sim 2$ milliarcsec (mas) is resolved into 3 different circular Gaussian model components of comparable full width at half-maximum (FWHM) sizes between $\sim 0.4-0.8$~mas. There are two nearly symmetrical components on both sides of the brightest, central one \cite{2018A&A...618A..68F}. Their
brightness temperatures are moderate ($T_\mathrm{b} \sim 10^{10}$~K), somewhat below the equipartition value of $T_\mathrm{b,eq} \approx 5 \times 10^{10}$~K \cite{1994ApJ...426...51R}. If $T_\mathrm{b,eq}$ is assumed as the intrinsic brightness temperature, this indicates that the Doppler factor $\delta=T_\mathrm{b} / T_\mathrm{b,eq} < 1$. In other words, there is no Doppler-boosted jet emission in J2102+6015 as would be expected if this was a blazar-type AGN with a relativistic jet pointing nearly to the line of sight.
Apart from the prominent resolved complex structure, a
weak mJy-level component appears at $\sim 10$~mas west of the brightness peak at 8.6~GHz \cite{2018A&A...618A..68F}. 

Another sensitive imaging experiment was performed with the Very Long Baseline Array (VLBA) at 8.4~GHz in 2017 February \cite{2021MNRAS.507.3736Z,2022ApJ...937...19Z}. The aim of that project was a jet kinematic study for a sample of nine high-redshift ($z>3.5$) radio quasars with well-identified mas-scale jet components and sufficiently long history of $X$-band VLBI imaging observations \cite{2022ApJ...937...19Z}. The source J2102+6015 was included in the sample. For the first time, its 8.4-GHz image revealed a faint emission feature located between the bright resolved complex structure in the east and the weak mJy-level component in the west \cite{2021MNRAS.507.3736Z}. 

The broad-band radio spectrum of J2102+6015 \cite{2017MNRAS.467.2039C,2021MNRAS.507.3736Z} is typical of gigahertz peaked-spectrum (GPS) sources, with $\sim 1$~GHz observed-frame turnover frequency. It corresponds to nearly 6~GHz in the source rest frame at the assumed redshift. Notably, all other very powerful ($P > 10^{28}$~W\,Hz$^{-1}$) VLBI-imaged high-redshift AGN (J0324$-$2918, J0906+6930, and J1606+3124 \cite{2016MNRAS.463.3260C}) also show peaked spectra \cite{2017MNRAS.467.2039C}, suggesting that they are young radio sources with jets currently piercing through the
dense interstellar medium \cite{2020NatCo..11..143A}.
For J2102+6015, the comparison of the sum of the VLBI component flux densities and the total flux density measurements at similar frequencies indicate that any radio emission extended beyond $\sim 10$~mas angular scales is insignificant in this source \cite{2017MNRAS.467.2039C,2021MNRAS.507.3736Z}. 

Concerning the kinematic properties of the mas-scale structure, the apparent separation between the bright eastern and the fainter western features remained practically unchanged over a period of $\sim 13$~yr, placing an upper limit of $0.04$~mas\,yr$^{-1}$ on the separation speed \cite{2021MNRAS.507.3736Z,2022ApJ...937...19Z}. The slow expansion, the peaked radio spectrum, the stable flux density on a long term, and the first-time detection of a (possibly) central component in between the previously known eastern and western emission features led \cite{2021MNRAS.507.3736Z} to propose a scenario where J2102+6015 is a young compact symmetric object (CSO). 

Most recently, an $S/X$-band astrometric VLBI study was devoted to measuring the absolute apparent proper motions of a sample of bright high-redshift radio quasars, including J2102+6015 \cite{Titov-AJ}. Interestingly, the densely-sampled position time series covering a time range of about 4.5~yr from 2017 to 2021 hint on a quasi-periodic variation, which is especially prominent in right ascension with an amplitude of $\sim 0.3$~mas and a period of about 3~yr. Certainly, longer-term observations that cover two or more suspected periods would be required to verify if the periodic signal persists \cite{Titov-AJ}. Nevertheless, assuming that the periodicity in the absolute astrometric position is real, one can speculate about its astrophysical origin. For example, a binary supermassive black hole system with sub-pc separation could produce jet precession that mimics the periodic change in the position of the `centroid' of the radio emission \cite{Titov-AJ}, presumably inside the brightest eastern emission feature.

The quasar J2102+6015 has never been studied with VLBI at frequency bands different from 2 and 8~GHz. Motivated by the apparently non-blazar nature and the puzzling mas-scale radio structure of this source, we conducted 5- and 22-GHz imaging experiments with the European VLBI Network (EVN) in 2021 June. Here we briefly describe the observations and data reduction, and show preliminary images in total intensity. The thorough analysis and interpretation of the polarization-sensitive EVN observations will be presented in a forthcoming publication (S. Frey et al., in preparation). 

\section{New EVN Observations and Data Analysis}

The EVN observations were done on 2021 June 7--8 at 22~GHz (project code: EF029A) and on 2021 June 14--15 at 5~GHz (EF029B). The total observing time was 8~h in both segments. At 22~GHz, the participating radio telescopes were Effelsberg (Ef, Germany), Medicina (Mc, Italy), Noto (Nt, Italy), Onsala 20-m telescope (O6, Sweden), Tianma (T6, China), Urumqi (Ur, China), Yebes (Ys, Spain), Svetloe (Sv, Russia), Zelenchukskaya (Zc, Russia), Badary (Bd, Russia), Sardinia (Sr, Italy), and Mets\"ahovi (Mh, Finland), supplemented by 3 antennas of the Korean VLBI Network (KVN), Yonsei (Ky), Ulsan (Ku), and Tamna (Kt). At 5~GHz, the network consisted of Ef, Mc, T6, Ur, Ys, Sv, Zc, Bd, Jodrell Bank Mk2 (Jb, United Kingdom), Westerbork (Wb, The Netherlands), Toru\'n (Tr, Poland), and Irbene (Ir, Latvia). Apart from the target source, the calibrators   J2055+6122 (for phase-referencing scans), J1407+2827, J0319+4130 (for polarization D-term calibration), and the fringe-finder J2022+6136 were occasionally also observed. The data were recorded in 8 intermediate frequency (IF) channels, each having 64 spectral points, in left and right circular polarizations. The data rate was 2~Gbit\,s$^{-1}$. The data were correlated at the Joint Institute for VLBI ERIC (Dwingeloo, The Netherlands) with 2~s integration time.

The visibility data were edited and calibrated in the US National Radio Astronomy Observatory (NRAO) Astronomical Image Processing System (\textsc{aips}) software \cite{2003ASSL..285..109G} in a standard way. A detailed account of the data reduction steps will be given elsewhere (S. Frey et al., in preparation). At 22~GHz, due to the resolved nature of the source, it was difficult to find interferometric fringes on long baselines between European and East-Asian antennas. Therefore we divided the network into two subarrays, scattered around the most sensitive radio telescopes in Europe (Ef) and in Asia (T6) as respective reference antennas. After successful fringe-fitting in both subarrays, the two calibrated visibility data sets were combined for imaging. Hybrid mapping was performed in the Caltech \textsc{difmap} package \cite{1994BAAS...26..987S}.

\section{Results and Discussion}

\begin{figure}
\centering
\includegraphics[width=0.95\linewidth]{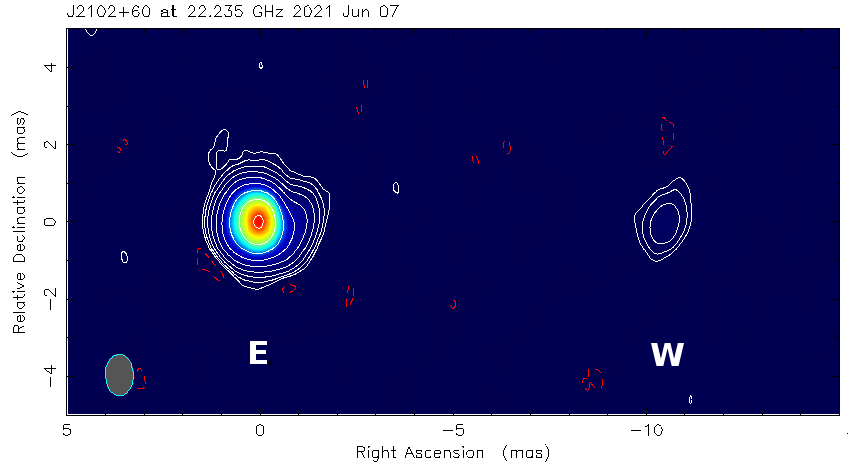}
\caption{Naturally weighted 22-GHz EVN image of J2102+6015. The peak brightness is 29.4~mJy\,beam$^{-1}$. The lowest contours are at $\pm 109~\mu$Jy\,beam$^{-1}$ ($3\sigma$ noise). The positive contour levels increase by a factor of 2. The elliptical Gaussian restoring beam (half-power width $1.08~\mathrm{mas} \times 0.72~\mathrm{mas}$, major axis position angle $1.4^{\circ}$) is shown in the lower left corner. The brighter eastern and the fainter western features are labeled as E and W, respectively.}\label{fig:kband}
\end{figure}

The mas-resolution 22-GHz EVN image of J2102+6015 (Fig.~\ref{fig:kband}) well reproduces the two main structural elements, the brighter eastern (E) and the fainter western (W) features known from the earlier 8-GHz images of the source \cite{2002ApJS..141...13B,2008AJ....136..580P,2018A&A...618A..68F,2021MNRAS.507.3736Z,2022ApJ...937...19Z}. Their separation is about 10~mas. The eastern complex is clearly resolved and extended towards the west, consistently with the brightness distribution models obtained at 8~GHz \cite{2018A&A...618A..68F,2021MNRAS.507.3736Z}. However, there is no trace of the faint `central' component found at 8.4~GHz using the most sensitive VLBA observations in 2017 \cite{2021MNRAS.507.3736Z}. At the expected location (4.7~mas from the brightness peak), the image brightness does not reach the 3-$\sigma$ noise level of $\sim 0.1$~mJy\,beam$^{-1}$. For comparison, the 8.4-GHz component had $1.3 \pm 0.2$~mJy\,beam$^{-1}$ peak brightness in 2017 \cite{2021MNRAS.507.3736Z}. The absence of this component could be due to its time variability or its steep spectrum. Note that at $z=4.575$, this observing frequency corresponds to $(1+z)$ times higher value, 124~GHz, in the rest frame of the source. 

\begin{figure}
\centering
\includegraphics[width=0.95\linewidth]{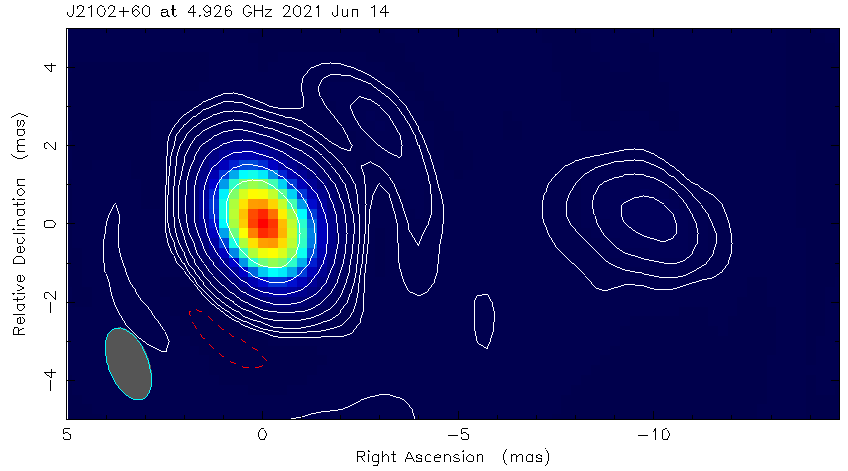}
\caption{Naturally weighted 5-GHz EVN image of J2102+6015. The peak brightness is 124~mJy\,beam$^{-1}$. The lowest contours are at $\pm 200~\mu$Jy\,beam$^{-1}$ ($3\sigma$ noise). The positive contour levels increase by a factor of 2. The elliptical Gaussian restoring beam (half-power width $1.94~\mathrm{mas} \times 1.02~\mathrm{mas}$, major axis position angle $21.3^{\circ}$) is shown in the lower left corner.}\label{fig:cband}
\end{figure}

The 5-GHz EVN image (Fig.~\ref{fig:cband}) also shows the two major features, in the east and west. The angular scale is the same as in the 22-GHz image (Fig.~\ref{fig:kband}). At the redshift $z=4.575$, 1~mas angular distance corresponds to 6.553~pc projected linear distance in the $\Lambda$CDM cosmological model with parameters $H_0=70$~km\,s$^{-1}$\,Mpc$^{-1}$, $\Omega_\mathrm{m}=0.3$, and $\Omega_{\Lambda}=0.7$.
The `central' component seen at 8.4~GHz \cite{2021MNRAS.507.3736Z} is not convincingly present here either. There are low-level spurious features around the brightest component that might be reduced with more careful data flagging and imaging. The bright eastern emission is resolved, with a faint extension towards northwest. Notably, some 8-GHz VLBA images, especially the most sensitive one obtained in 2017 February, also hint on an extension in the same position angle \cite{2021MNRAS.507.3736Z}. 

The turnover frequency for an optically thick spectrum $\nu_t$ (see e.g. eq. 4.56 of \cite{2013LNP...873.....G}) lies at $\approx 10.9$~GHz (corresponding to an observer frame frequency of $\approx 2$~GHz) assuming an equipartition magnetic field strength of mG and an emitting region of size 0.1~pc, consistent with the observational inference of a spectral peak at $\approx 1$~GHz \cite{2021MNRAS.507.3736Z}. Our VLBI observations at 5 and 22~GHz are therefore in the optically thin emission regime, indicating a slowly growing jet in the CSO phase \cite{2007MNRAS.381.1548K,2012ApJ...760...77A}, consistent with the expectation from previous studies. This then implies that the radio power, governed by the conversion of jet kinetic to radiative energy will continue to increase with time with the source expansion.

\section{Summary}

The radio source J2102+6015 is still puzzling in many ways. Its high spectroscopic redshift ($z=4.575$) is a tentative measurement and would need independent confirmation. Recent monitoring of its absolute astrometric position suggests a harmonic motion in the sky plane with a period of $\sim 3$~yr, but the observations cover an interval only 1.5 times longer than the suspected period \cite{Titov-AJ}. Earlier $S/X$-band VLBI imaging observations revealed a structure that is dominated by a bright mas-scale feature in the east. It is, however, resolved into more components when fitted with Gaussian brightness distribution models. The most sensitive 8.4-GHz VLBA image of J2102+6015 obtained to date \cite{2021MNRAS.507.3736Z} revealed a faint `central' component between the eastern and western features which may (or may not) be a core of a CSO.

Here we presented preliminary results and total intensity images at 5 and 22~GHz from our EVN experiments conducted in 2021 June. J2102+6015 was imaged with VLBI for the first time at these frequencies. The overall source structure is consistent with what is observed previously at 8~GHz, but we did not detect the elusive `central'  component. More detailed, quantitative analysis including polarization imaging, brightness distribution modeling, spectral index mapping is deferred to a forthcoming publication. 

\section*{Acknowledgements}
The European VLBI Network is a joint facility of independent European, African, Asian and North American radio astronomy institutes. The EVN observations presented in this paper have been conducted under the project code EF029. This research was supported by the Hungarian National Research, Development and Innovation Office (NKFIH), grant number OTKA K134213.


\begin{thebibliography}{99}

\bibitem[\protect\citeauthoryear{An \& Baan}{2012}]{2012ApJ...760...77A} An T., Baan W.~A., 2012, ApJ, 760, 77. doi:10.1088/0004-637X/760/1/77

\bibitem[\protect\citeauthoryear{An et al.}{2020}]{2020NatCo..11..143A} An T., Mohan P., Zhang Y., Frey S., Yang J., et al., 2020, Nat. Commun., 11, 143. doi:10.1038/s41467-019-14093-2
 
\bibitem[\protect\citeauthoryear{Beasley et al.}{2002}]{2002ApJS..141...13B} Beasley A.~J., Gordon D., Peck A.~B., Petrov L., MacMillan D.~S., et al., 2002, ApJS, 141, 13. doi:10.1086/339806

\bibitem[\protect\citeauthoryear{Chambers et al.}{2016}]{2016arXiv161205560C} Chambers K.~C., Magnier E.~A., Metcalfe N., Flewelling H.~A., Huber M.~E., et al., 2016, arXiv preprint 1612.05560, doi:10.48550/arXiv.1612.05560

\bibitem[\protect\citeauthoryear{Coppejans et al.}{2016}]{2016MNRAS.463.3260C} Coppejans R., Frey S., Cseh D., M\"uller C., Paragi Z., et al., 2016, MNRAS, 463, 3260. doi:10.1093/mnras/stw2236

\bibitem[\protect\citeauthoryear{Coppejans et al.}{2017}]{2017MNRAS.467.2039C} Coppejans R., van Velzen S., Intema H.~T., M\"uller C., Frey S., et al., 2016, MNRAS, 467, 2039. doi:10.1093/mnras/stx215

\bibitem[\protect\citeauthoryear{Frey et al.}{2018}]{2018A&A...618A..68F} Frey S., Titov O., Melnikov A.~E., de Vicente P., Shu F., 2018, A\&A, 618, A68. doi:10.1051/0004-6361/201832771

\bibitem[\protect\citeauthoryear{Ghisellini}{2013}]{2013LNP...873.....G} Ghisellini G., 2013, Radiative Processes in High Energy Astrophysics, Lecture Notes in Physics (Cham: Springer), 873. doi:10.1007/978-3-319-00612-3

\bibitem[\protect\citeauthoryear{Greisen}{2003}]{2003ASSL..285..109G} Greisen E.~W., 2003, in Information Handling in Astronomy -- Historical Vistas, ed. A. Heck, Astrophysics and Space Science Library (Dordrecht: Kluwer), 285, 109. doi:10.1007/0-306-48080-8\_7

\bibitem[\protect\citeauthoryear{Kaiser \& Best}{2007}]{2007MNRAS.381.1548K} Kaiser C.~R., Best P.~N., 2007, MNRAS, 381, 1548. doi:10.1111/j.1365-2966.2007.12350.x

\bibitem[\protect\citeauthoryear{Petrov et al.}{2008}]{2008AJ....136..580P} Petrov L., Kovalev Y.~Y., Fomalont E.~B., Gordon D., 2008, AJ, 136, 580. doi:10.1088/0004-6256/136/2/580

\bibitem[\protect\citeauthoryear{Readhead}{1994}]{1994ApJ...426...51R} Readhead A.~C.~S., 1994, ApJ, 426, 51. doi:10.1086/174038

\bibitem[\protect\citeauthoryear{Shepherd, Pearson, \& Taylor}{1994}]{1994BAAS...26..987S} Shepherd M.~C., Pearson T.~J., Taylor G.~B., 1994, BAAS, 26, 987

\bibitem[\protect\citeauthoryear{Snios et al.}{2020}]{2020ApJ...899..127S} Snios B., Siemiginowska A., Sobolewska M., Cheung C.~C., Kashyap V., et al., 2020, ApJ, 899, 127. doi:10.3847/1538-4357/aba2ca

\bibitem[\protect\citeauthoryear{Sowards-Emmmerd et al.}{2004}]{2004ApJ...609..564S} Sowards-Emmerd D., Romani R.~W., Michelson P.~F., Ulvestad J.~S., 2004, ApJ, 609, 564. doi:10.1086/421239
 
\bibitem[\protect\citeauthoryear{Titov et al.}{2023}]{Titov-AJ} Titov O., Frey S., Melnikov A., Shu F., Xia B., et al., 2023, AJ, in press 

\bibitem[\protect\citeauthoryear{Zhang et al.}{2021}]{2021MNRAS.507.3736Z} Zhang Y., An T., Frey S., Yang X., Krezinger M., et al., 2021, MNRAS, 507, 3736. doi:10.1093/mnras/stab2289

\bibitem[\protect\citeauthoryear{Zhang et al.}{2022}]{2022ApJ...937...19Z} Zhang Y., An T., Frey S., Gab\'anyi K.~\'E., Sotnikova Y., 2022, ApJ, 937, 19. doi:10.3847/1538-4357/ac87f8

\end{thebibliography}
\end{document}